\documentclass{article}

\usepackage[final]{nips_2016}

\usepackage[utf8]{inputenc} 
\usepackage[T1]{fontenc}    
\usepackage{hyperref}       
\usepackage{url}            
\usepackage{booktabs}       
\usepackage{amsfonts}       
\usepackage{nicefrac}       
\usepackage{microtype}      
\usepackage{amsmath,amssymb}
\usepackage{graphicx}

\title{Demographical Priors for Health Conditions Diagnosis Using Medicare Data}

  \author{
   Fahad Alhasoun \\
   Center for Computational  \\Engineering\\
   MIT\\
   \texttt{fha@mit.edu} \\
  \And
  May Alhazzani \\
   Center for Complex\\ Engineering Systems at \\KACST and MIT \\
   \texttt{mayh@mit.edu} \\
   \And
   Marta C. Gonz\'alez \\
   Department of Civil and\\Environmental Engineering \\
   MIT\\
   \texttt{martag@mit.edu} \\
  }

\begin{document}

\maketitle

\begin{abstract} This paper presents an example of how demographical characteristics of patients influence their susceptibility to certain medical conditions.  In this paper, we investigate the association of health conditions to age of patients in a heterogeneous population. We show that besides the symptoms a patients is having, the age has the potential of aiding the diagnostic process in hospitals. Working with Electronic Health Records (EHR), we show that medical conditions group into clusters that share distinctive population age densities. We use Electronic Health Records from Brazil for a period of 15 months from March of 2013 to July of 2014. The number of patients in the data is 1.7 million patients and the number of records is 47 million records. The findings have the potential of helping in a setting where an automated system undergoes the task of predicting the condition of a patient given their symptoms and demographical information. 
\end{abstract}

\section{Introduction}
Studies of comorbidities and diseases correlations have been mainly focusing on few diseases using techniques of hypothesis-testing \cite{camilo2004seizures,murray2012illness,murtagh2011trajectories,murtagh2008illness,teno2001dying,finkelstein2009chronic} or with focus on certain co-morbidities to index diseases\cite{petri2010data}. Recently, researchers started studying massive health records to uncover the associations and patterns in complex diseases demonstrating the potentials of studying health records \cite{jensen2012mining,blair2013nondegenerate,murtagh2011trajectories,chen2009cancer}. However, medical conditions data originate from various sources have a narrowed set of the general population of patients.  Studies used statistical techniques to produce fine grained patient stratification and disease co-occurence statistic of patients in the Sct. Hans Hospital (the largest Danish psychiatric hospital) \cite{roque2011using}. Studies used a network approach to analyze data covering 3 years of medical care claims, of patients who are 65 years or older, which biased the analysis towards the population of the elderly\cite{hidalgo2009dynamic}. However, recently studies were conducted on data of more heterogeneous populations. The studies focused on uncovering patterns centered on a small number of key diagnoses to detect diseases earlier in a patients life \cite{jensen2014temporal}. Another study analyzed the structure of co-morbidity networks on five predefined age intervals of patients  \cite{chmiel2014spreading}.
\\ 
The wisdom of doctors when it comes to assessing the susceptibility to medical conditions have been influenced by the years of practice and observation of many cases on daily basis. Doctors' knowledge of the susceptibility of diseases to different ages/genders serves as an essential prior to perform diagnostics of incoming patients. Similar symptoms for patients might lead to different diagnosis depending on the age and gender of the patient, a patient who is 70 years old is more probable to be suffering a heart attack than that of a 10 years old even if both patients are suffering the symptom of chest pain.  We show here that besides the symptoms a patient is having, age has the potential of significantly aiding the diagnostic process. In this paper, we aim at uncovering the relationship between health conditions and the age of a patient. We stratify health conditions that share similar population age densities.

\section{Age Densities Signatures}
\begin{figure}[h!]
\centering
  \includegraphics[width=0.9\textwidth]{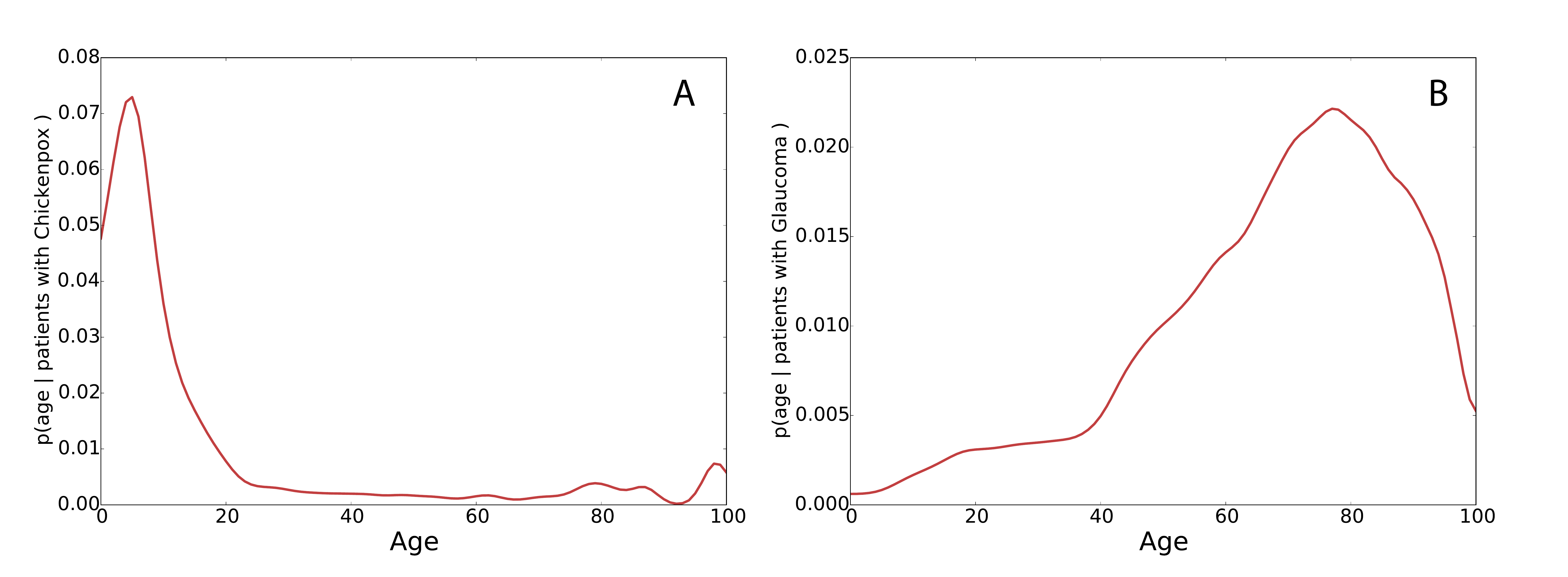}    
\caption{Kernel density estimation of the probability density function of the age of patients with Chickenpox (A) and Glaucoma (B)}
\label{example_dists}
\end{figure}

The data used in the paper pertains to records of health insurance claims from Brazil for a period of 15 months from March of 2013 to July of 2014. The  number of patients in the data is around 1.7 million  patients. For each patient in the data, a log for each visit to the doctor is stored in the database. The database uses the International Classification of Diseases version 10 (ICD-10) \cite{world2004international}.  ICD-10 has a range of 23k codes each representing a health condition. The data includes the demographical attributes of the population (i.e.  age and gender information of the patient). The total number of visit records in the data is about 6.6 million records.\\ By inspection, ICD-10 codes have distinctive signatures of density on the age dimension that spans the various age groups from birth onward. Figure \ref{example_dists} shows example age density signatures of Chickenpox and Glaucoma. As expected, Glaucoma is more prevalent for the older age groups \cite{centers2011state} while Chickenpox is more prevalent in younger age groups\cite{kliegman2016nelson,centers2011epidemiology}.  The examples of the distributions hint that there is a pattern of higher likelihood of health conditions for certain age groups in contrast to others. The rest of the paper aims to investigate this further.

\section{Hierarchal  Clustering of Health Conditions}
To uncover common patterns of the age distribution of ICD-10 codes, we used a Hierarchical Agglomeration Clustering (HAC) approach to group the codes based on the similarities of age distributions. Each code is represented by a vector $v$ of length 100 where each cell represents $p(age=i|patients \in C)$ where $patients \in C$  is the set of patients with the condition on their records. The clusters are decided as the ones minimizing the variance of distances between codes in the cluster while maximizing the distance between the clusters (i.e. the ward minimum variance method) \cite{ward1963hierarchical}.
\\
 HAC cluster vectors, where each vector is a representation of the probability mass function of a code in the data.  The vector representation of the probability mass function of the ages of a ICD-10 code is as follows:
\begin{equation}
p(age| patient \in code)=[ p_{1}, p_{1} , ...., p_{100}]
\end{equation}
Where $p_{i}=p(age=i| patient \in code)$ for a given code.  At initialization, HAC assigns each vector object to a cluster, and sequentially merging them into clusters until all codes form one cluster. For measuring the distance  $d$  between two vector representations of age density, we use euclidean distance. The Ward distance criterion of clusters is dependent on the within cluster distances  and the across clusters distances.   Ward algorithm is conservative when merging clusters, thus it tends to find very compact clusters \cite{ward1963hierarchical}.


   
HAC provides a hierarchy structure of the clustered codes as illustrated in figure \ref{clusters}. To determine the number of clusters $k$ that best divide the data, we calculate the total within-cluster distances for  $k$ from 1 to 20. The variance ratio drops as $k$ increase until it does not decrease significantly. We select  $k$ that corresponds to the point where the total distances stops decreasing significantly. This method is known as the elbow curve method.

\section{Results}
\begin{figure}[t]
\includegraphics[width=1\textwidth]{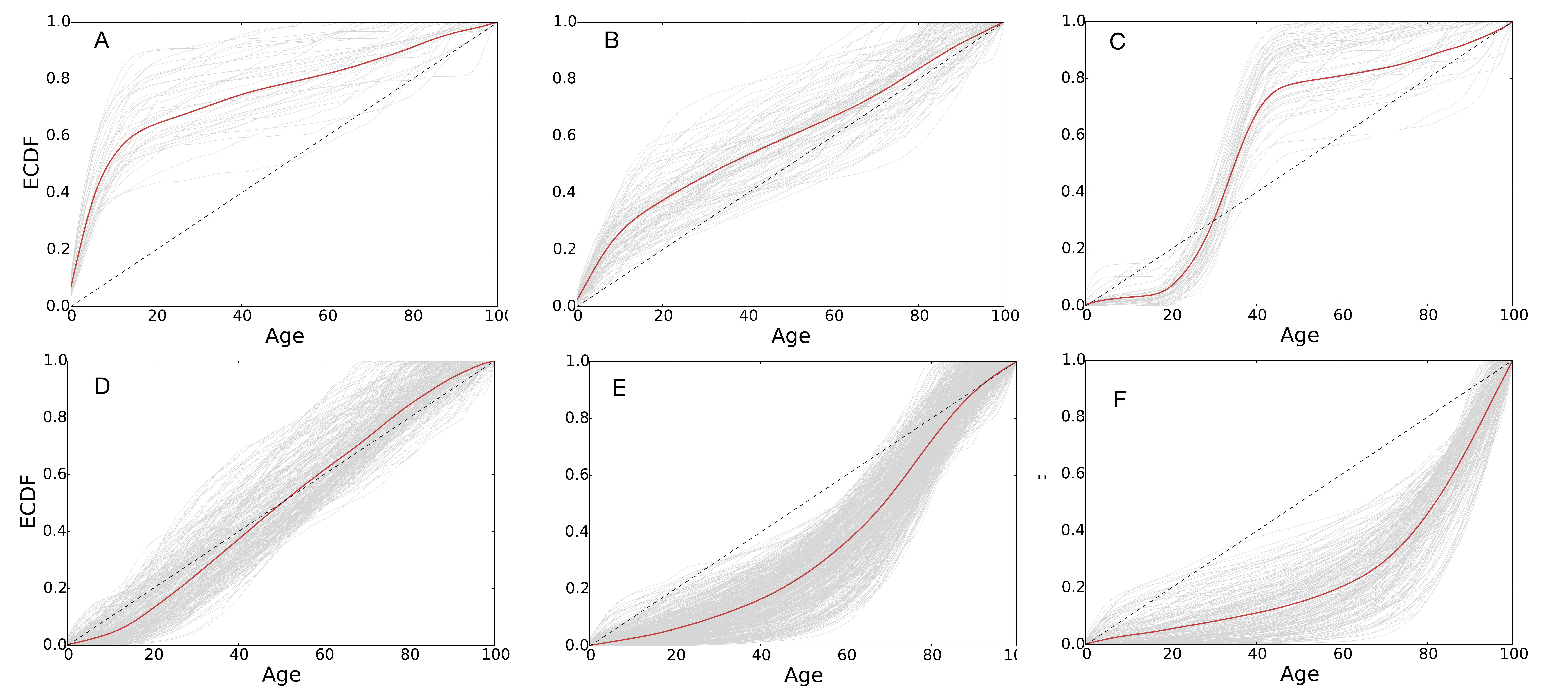}    
\caption{ ECDF in gray represent cumulative distribution of $P(age|patients \in cpde)$ and ECDF in red are the cluster averages for illustration. The clusters of ICD-10 codes given by the HAC are label by the alphabet A to F. Cluster A of ICD-10 codes have more concentration towards the infants and children. Cluster B of conditions having a density around an increased age interval than that of A that is typically more concentrated around teenage years and early adulthood. Cluster C has a the narrowest concentration of age in the thirties. Cluster D is for codes that are closest to being uniform in adulthood. Cluster F of codes have higher concentration in the ages after the 60s. Cluster D of ICD-10 codes are patients mostly in the ages 80 and above.}
\label{clusters}
\end{figure}
\begin{figure}[t!]

  \includegraphics[width=1\textwidth]{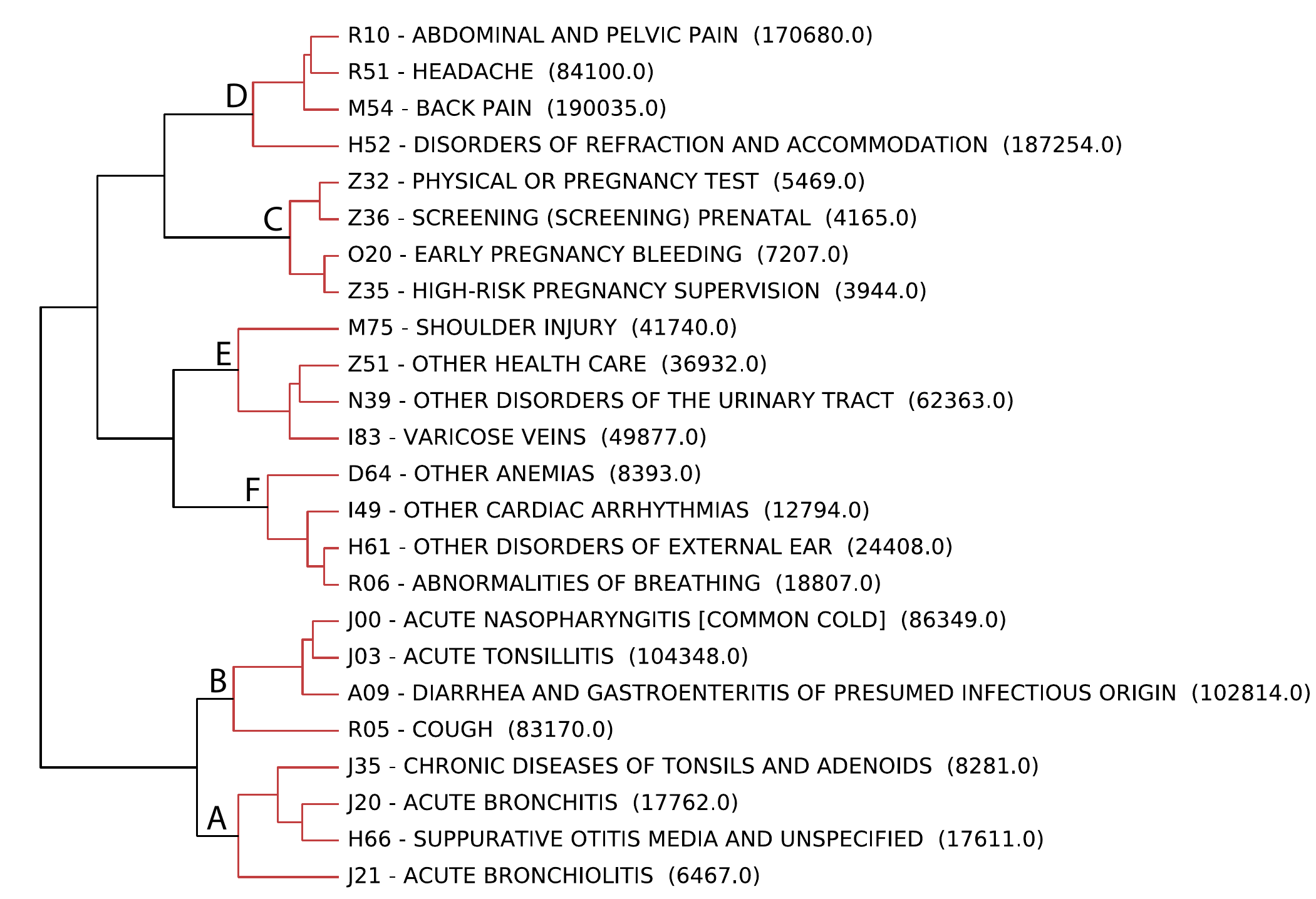}    
\caption{Hierarchical clustering of the four most frequent ICD-10 codes for each cluster, the frequency of each ICD-10 code is in parenthesis. The alphabet letters assignments correspond to the clusters discussed in figure \ref{clusters}.}
\label{hirarchy}
\end{figure}

 The age distribution of the codes clusters into six main clusters as shown in Figure \ref{clusters}.
Clusters A and B show two clusters of codes having higher density towards the lower spectrum of age. Cluster C shows a group of codes that have age densities concentrated in the ages 20 to 40 years old. Cluster D has conditions that are almost uniformly distributed across the ages. Cluster E had codes with densities concentrating in the range of ages after 60 and cluster F had codes with age densities concentrating after the ages of seventies years old.

figure \ref{hirarchy} illustrates a few examples of ICD-10 codes from the clusters discovered in the data. For each cluster, figure \ref{hirarchy} shows four example conditions. The branches are labeled by their clusters from A to F. The example codes are the highest four in terms of their frequency in occurrence in the data. Within cluster A, J20 Acute Bronchitis and H66 Suppurative Otitis were observed in 17 thousand patients each and both have a concentration towards the lower ages as shown. Cluster B has A09 Diarrhea and J03 Acute Tonsillitis both with around 100 thousand patients. The noticed increase of the number of patients is due to the non-uniform population age distribution where the majority of the population is in the age range of early and middle adulthood. Cluster C has the conditions Z35 Early Pregnancy Bleeding with around 7 thousand patients and Z32 Physical and Pregnancy Test with 5 thousands patients. In general, we find that conditions in cluster C pertain to pregnancy, childbirth and postpartum conditions explaining the concentrated age distribution in figure \ref{clusters}-C. Cluster D has M54 Back Pain with 190 thousands patients and R10 Abdominal and Pelvic Pain with 170 thousand patients. We notice that as the clusters have more density around the peak of the age distribution of the population, the number of patients per code in the clusters becomes relatively high. Cluster E with age density more towards the elderly has R08 Abnormalities of Breathing as the most common with around 19 thousands patients and I50 Head Injuries in 8 thousand patients, also notice that Heart Failure comes third with around 7 thousands patients. Cluster F with age density well into the elderly age has M75 Shoulder Injury with 41 thousands patients and N40 Prostatic Hyperplasia with around 28 thousands patients. 

\section{Conclusion}
In this paper, we show an example where demographical characteristics of patients namely the age of a patient could potentially aid the diagnostic process. Other demographical characteristics such as gender, race, income level among others have the potential in aiding the diagnostic process for incoming patients. The goal in this paper is to help build such age susceptibility prior knowledge for automated diagnostics in a setting where an automated system goes through the task of predicting the condition of a patient given their symptoms. Upon the availability of data, the study of the association of demographical characteristics such as gender, race and income level has the potential of uncovering useful prior knowledge to encode in models for predicting medical conditions given symptoms and demographical characteristics.

\bibliographystyle{plain}	
\bibliography{citation.bib}

\end{document}